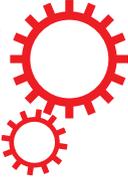

OPEN

# Time-reversal invariant resonant backscattering on a topological insulator surface driven by a time-periodic gate voltage

Ming-Xun Deng[1,2], R. Ma[3], Wei Luo[4], R. Shen[1,5], L. Sheng[1,5] & D. Y. Xing[1,5]

We study the scattering of the Dirac electrons by a point-like nonmagnetic impurity on the surface of a topological insulator, driven by a time-periodic gate voltage. It is found that, due to the doublet degenerate crossing points of different Floquet sidebands, resonant backscattering can happen for the surface electrons, even without breaking the time-reversal (TR) symmetry of the topological surface states (TSSs). The energy spectrum is reshuffled in a way quite different from that for the circularly polarized light, so that new features are exhibited in the Friedel oscillations of the local charge and spin density of states. Although the electron scattering is dramatically modified by the driving voltage, the $1/\rho$ scale law of the spin precession persists for the TSSs. The TR invariant backscattering provides a possible way to engineer the Dirac electronic spectrum of the TSSs, without destroying the unique property of spin-momentum interlocking of the TSSs.

Topological insulators (TIs), characterized by the topologically protected gapless boundary states, have enjoyed a surge of research interest in condensed-matter physics[1–8]. The gapless boundary states inside the bulk band gap distinguish the TIs from ordinary insulators, adiabatically. In a three-dimensional TI, the boundary states are two-dimensional topological surface states (TSSs), in which the spins of electrons are locked to their momenta in a chiral structure[2–4,8,9]. As a consequence, the gapless TSSs are robust to static time-reversal (TR) invariant perturbations, and perfect backscattering from TR invariant impurities is prohibited. The unique topological properties make TIs promising materials for applications in spintronics and topological quantum computation[10,11], in which manipulation of the Dirac electronic properties or engineering the Dirac spectrum of the TSSs is often needed[12–14]. Artificially creating an energy gap in the spectrum of the TSSs by breaking the TR symmetry could induce topological transitions of the TIs, and thus is thought to be one of the most promising pathways to manipulate the Dirac electrons[15–22].

Recently, application of a time-periodic perturbation to the TIs attracted much attention[23–25]. The time-periodic driving, by reshuffling the spectrum, can also induce topological transitions of the TIs. For example, a time-periodic driving can split the static spectrum into Floquet sidebands[24,26,27], accompanied with many interesting physical phenomena unique to the periodically driven systems, such as the anomalous edge states[24,28], Majorana $\pi$ modes[29,30] and Floquet multi-Weyl points[31]. Due to the reshuffled spectrum, new scattering channels will be opened up for the Dirac electrons, and the quasiparticle interference also displays new interesting characteristics. As discussed in ref.[32], the Dirac electron backscattering is allowed even for a nonmagnetic impurity, when the TI surface is irradiated by a beam of circularly polarized light. However, the circularly polarized light, which is widely used to induce topological transitions in periodically driven systems[33–36], is known to break the TR symmetry. Therefore, the occurrence of perfect backscattering on the TI surface is naturally attributable to the TR symmetry breaking of the TSSs caused by the circularly polarized light.

[1]National Laboratory of Solid State Microstructures and Department of Physics, Nanjing University, Nanjing, 210093, China. [2]Laboratory of Quantum Engineering and Quantum Materials, ICMP and SPTE, South China Normal University, Guangzhou, 510006, China. [3]Jiangsu Key Laboratory for Optoelectronic Detection of Atmosphere and Ocean, Nanjing University of Information Science and Technology, Nanjing, 210044, China. [4]School of Science, Jiangxi University of Science and Technology, Ganzhou, 341000, China. [5]Collaborative Innovation Center of Advanced Microstructures, Nanjing University, Nanjing, 210093, China. Correspondence and requests for materials should be addressed to L.S. (email: shengli@nju.edu.cn)

The backscattering induced by TR symmetry breaking has been investigated extensively both for static systems, e.g., magnetically doped TI surfaces[13,37–40], topological superconductors[41,42], vibrational defects[43], Weyl semimetals[44,45] and circularly polarized light driven TI surfaces[32]. However, when the TR symmetry is broken, accompanying the gap opening effect, the spin-momentum interlocking of the TSSs could also be destroyed, which may impair the applications taking advantage of this unique feature of the TSSs in spintronics and topological quantum computation. Therefore, it is desirable to find a way to engineer the TSSs, without destroying the spin-momentum interlocking of the Dirac fermions. Interestingly, as we will show in this paper, in the presence of a time-periodic gate voltage, perfect backscattering can happen for the TSSs, in the absence of any TR breaking perturbations, which presents a new mechanism to manipulate the surface electrons.

Motivated by the advances in the Floquet TIs, especially the experimental realization of the Floquet bands in photonic crystals[46] and on TI surfaces[47,48], there is increasing interest focused on detailed study of the quantum phenomena in periodically driven systems. In this paper, we consider a TI surface driven by a time-periodic gate voltage, and investigate the characteristics of impurity scattering of the Dirac fermions on the TI surface. Although the gate voltage does not break the TR symmetry for the TI surface, it is interesting to find that perfect backscattering becomes allowable for a TR invariant impurity, which gives rise to some new features of electron scattering for the TSSs. The energy spectrum, in the presence of the gate voltage, is split into multi-sidebands, in a way quite different from that for the circularly polarized light. In the present system, the linear dispersion and Dirac point still remain in the energy spectrum, while for the circularly polarized light, the linear dispersion is dramatically modified and the Dirac point disappears. Consequently, new features are exhibited in the Friedel oscillations of the local charge and spin density of states (DOS). Although the electron scattering is dramatically modified by the gate voltage, the $1/\rho$ scale law of the spin procession remains for the gate-voltage-driven TSSs.

## Results

**Model and theory.** Let us consider a TI surface driven by a time-periodic external field, which is described by the Hamiltonian $H = \sum_{\mathbf{k}} c_{\mathbf{k}}^{\dagger} \mathcal{H}(\mathbf{k}, t) c_{\mathbf{k}}$ with $\mathcal{H}(\mathbf{k}, t) = \mathcal{H}(\mathbf{k}) + \mathcal{H}_{ext}(t)$. Here, $\mathcal{H}(\mathbf{k}) = \hbar v_F \boldsymbol{\sigma} \cdot (\hat{z} \times \mathbf{k})$ is a low-energy effective Hamiltonian for the TI surface, in which $v_F$, $\boldsymbol{\sigma} = (\sigma_x, \sigma_y)$ and $\hbar \mathbf{k} = (p_x, p_y)$ represent the Fermi velocity, vector of Pauli matrices and in-plane momenta for the surface electrons, respectively. The time-periodic external field is described by $\mathcal{H}_{ext}(t)$, which satisfies the time-periodic condition $\mathcal{H}_{ext}(t) = \mathcal{H}_{ext}(t + nT)$ with $T$ as the period of the external field. For simplicity, we consider that the external field has only a single frequency, i.e.,

$$\mathcal{H}_{ext}(t) = [\mu_0 + 2A_0 \cos(\Omega t)]\sigma_0 + e\mathbf{A}(t) \cdot \boldsymbol{\sigma} \tag{1}$$

with the term tied to $\sigma_0$ generated by a time-dependent gate voltage, where $A_0$ and $\Omega$ represent the amplitude and frequency, respectively, and $\mathbf{A}(t) = A_c[\lambda \sin(\Omega t), \cos(\Omega t)]$ denotes a circularly polarized light of amplitude $A_c$, frequency $\Omega$ and right/left polarization $\lambda = \pm 1$. Here, we include the circularly polarized light, which breaks the TR symmetry, in order to make comparison between the TR invariant and TR symmetry broken backscattering later. The circularly polarized light is widely used to induce topological phase transitions in periodically driven systems[23–25,33–36]. However, the result is natural, as the TR symmetry is broken by the circularly polarized light, perfect backscattering can in principle occur on the TI surface. Interestingly, as we will show, perfect backscattering can happen, in the absence of any TR symmetry breaking perturbations, for the TI surface states driven by a time-periodic voltage.

An impurity absorbed on the TI surface is described by a potential energy of the general form[13,18,22,32,39,41], $U = \int d\mathbf{r} c^{\dagger}(\mathbf{r}) V \delta(\mathbf{r} - \mathbf{r}_{imp}) c(\mathbf{r})$, where $c^{\dagger}(\mathbf{r}) = \frac{1}{\sqrt{N}} \sum_{\mathbf{k}} c_{\mathbf{k}}^{\dagger} e^{-i\mathbf{k} \cdot \mathbf{r}}$ is the electron creation operator at position $\mathbf{r}$, and the $\delta$ function is used to approximate a spatially continuous scattering potential sharply peaked at the impurity site $\mathbf{r}_{imp}$. Without loss of generality, we assume that the impurity potential energy comprises of a charge potential of strength $U_0$ and a spin-dependent magnetic potential of strength $U_M$, namely, $V = U_0 \sigma_0 + U_M \mathbf{s} \cdot \boldsymbol{\sigma}$. Here, we in general include the spin-dependent potential, which breaks the TR symmetry, mainly for comparison between the TR invariant and TR broken backscattering. Consequently, the total Hamiltonian for the system is given by

$$H_{tot} = \sum_{\mathbf{k}} c_{\mathbf{k}}^{\dagger} \left[ \mathcal{H}(\mathbf{k}, t) c_{\mathbf{k}} + \sum_{\mathbf{k}'} U_{\mathbf{k}\mathbf{k}'} c_{\mathbf{k}'} \right], \tag{2}$$

where $U_{\mathbf{kk}'} = \frac{1}{N} V e^{-i(\mathbf{k} - \mathbf{k}') \cdot \mathbf{r}_{imp}}$ couples electron states of different momenta, $\mathbf{k}$ and $\mathbf{k}'$, due to the broken translational symmetry.

The information of the electron scattering with the impurity on the periodically driven TI surface can be extracted from the retarded Green's function, which is defined as

$$G_{\mathbf{k},\mathbf{k}'}(t, t') = -\frac{i}{\hbar} \theta(t - t') \langle \{c_{\mathbf{k}}(t), c_{\mathbf{k}'}^{\dagger}(t')\} \rangle \tag{3}$$

with $\theta(x)$ being the heaviside function. For single frequency driving, the impurity-perturbed Green's function is given by[32]

$$\mathcal{G}_{\mathbf{k},\mathbf{k}'}^{m,n}(\varepsilon) = \mathcal{G}_{\mathbf{k}}^{m,n}(\varepsilon) \delta_{\mathbf{kk}'} + \sum_{pq} \mathcal{G}_{\mathbf{k}}^{m,p}(\varepsilon) T_{\mathbf{kk}'}^{p,q}(\varepsilon) \mathcal{G}_{\mathbf{k}'}^{q,n}(\varepsilon), \tag{4}$$

where $\mathcal{G}_{\mathbf{k},\mathbf{k}'}^{m,n}(\varepsilon)$ is the Fourier transform of the Floquet Green's function $G_{\mathbf{k},\mathbf{k}'}(t, t', \varepsilon)$ and

$$T^{p,q}_{\mathbf{k}\mathbf{k}'}(\varepsilon) = U_{\mathbf{k}\mathbf{k}'}\left[\delta_{pq} - \frac{1}{N}\sum_{\mathbf{q}} G^{-p,-q}_{\mathbf{q}}(\varepsilon)V\right]^{-1} \quad (5)$$

is the T matrix. In the specific calculation, we will replace the summation by an integral $\frac{1}{N}\sum_{\mathbf{q}} \to \frac{1}{(2\pi)^2}\int_{-k_c}^{k_c} d^2\mathbf{q}$, where $k_c = \Lambda/(2\hbar v_F)$ is an ultraviolet cutoff related to the bandwidth $\Lambda$. The photon-modulated Green's functions

$$\mathcal{G}^{m,n}_{\mathbf{k}}(\varepsilon) = \mathcal{G}^{(m-n)}_{\mathbf{k}}(\varepsilon + n\hbar\Omega),$$

describing the energy absorption and emission processes, are determined by

$$\mathcal{G}^{(0)}_{\mathbf{k}}(\varepsilon) = \frac{1}{g_0^{-1} - \mathcal{H}_{-1}\mathcal{F}_1^+ - \mathcal{H}_{+1}\mathcal{F}_1^-} \quad (6)$$

and, for an arbitrary positive integer $n$,

$$\mathcal{G}^{(\pm n)}_{\mathbf{k}}(\varepsilon) = \prod_{m=1}^{n} \mathcal{F}^{\pm}_m \mathcal{G}^{(0)}_{\mathbf{k}}(\varepsilon), \quad (7)$$

where $\mathcal{F}^{\pm}_n$ are given by the iteration equations

$$\mathcal{F}^{\pm}_n = \frac{1}{g_{\pm n}^{-1} - \mathcal{H}_{\mp 1}\mathcal{F}^{\pm}_{n+1}}\mathcal{H}_{\pm 1} \quad (8)$$

with $\mathcal{H}_n = \frac{1}{T}\int_0^T \mathcal{H}(\mathbf{k}, t)e^{in\Omega t}dt$ and $g_l = (\varepsilon + l\hbar\Omega - \mathcal{H}_0)^{-1}$. Consequently, the averaged DOS for a fixed quasienergy $\varepsilon$ is given by

$$\begin{aligned}\rho_{\mathbf{k},\mathbf{k}'}(\varepsilon) &= -\frac{1}{\pi}\frac{1}{T}\int_0^T dt\, \text{Im}\{\text{Tr}[G_{\mathbf{k},\mathbf{k}'}(t, t, \varepsilon)]\} \\ &= -\frac{1}{\pi}\sum_n \text{Im}\{\text{Tr}[\mathcal{G}^{n,n}_{\mathbf{k},\mathbf{k}'}(\varepsilon)]\}.\end{aligned} \quad (9)$$

The relevant information is contained in $\rho_{\mathbf{k},\mathbf{k}'}(\varepsilon)$, which is related to Eq. (4). For example, by inspecting the poles of the first term of $\mathcal{G}^{n,n}_{\mathbf{k},\mathbf{k}'}(\varepsilon)$, we can obtain the photon-dressed spectrum for the $n$-th Floquet band, and the impurity scattering effect on the photon-dressed Dirac fermions is determined by the second term of Eq. (4).

**Resonant backscattering.** As mentioned above, it is not a surprise for TR broken perturbations to induce perfect backscattering on the TI surface. Here, the resonant backscattering that we consider is driven by the gate voltage, i.e., $A_c = 0$. Therefore, the neighbor coupling between the Floquet bands are $\mathcal{H}_{\pm 1} = A_0\sigma_0$. As it shows, during the energy absorption and emission processes, the spins of the electrons are unchanged, so that the TR symmetry remains for the surface states. On the other hand, to avoid TR broken perturbations, a nonmagnetic impurity would be taken into account, i.e., $U_M = 0$. The nonforward scattering is governed by the second term of Eq. (4), where the impurity couples electron states of different momenta and Floquet bands, $|m, \mathbf{k}\rangle$ and $|n, \mathbf{k}'\rangle$, indirectly. The scattering rate between $\mathbf{k}$ and $\mathbf{k}'$ for the $n$-th Floquet band is characterized by the magnetic response function $M^{(n)}(\mathbf{k}, \mathbf{k}') = \left\|(M_x^{(n)}, M_y^{(n)}, M_z^{(n)})(\mathbf{k}, \mathbf{k}')\right\|$, with

$$M^{(n)}_{i=(x,y,z)}(\mathbf{k}, \mathbf{k}') = -\frac{1}{2\pi}\text{Im}\{\text{Tr}[\sigma_i\mathcal{G}^{n,n}_{\mathbf{k},\mathbf{k}'}(\varepsilon)]\}. \quad (10)$$

We would consider a relatively weak external field, so that the scattering processes are dominated by states $|n=0, \mathbf{k}\rangle$, and we mainly focus on the behavior of the central Floquet band in the following. For a static TI surface, the electron backscattering is prohibited for a nonmagnetic impurity, due to the spin-momentum interlocking of the surface states. This is illustrated in Fig. 1(a), where a dark gap appears on the bright ring for $A_0 = 0$, indicating the absence of perfect backscattering. As we increase the amplitude of the gate voltage, e.g., $A_0 = 0.2$, 0.3, 0.5 in Fig. 1(b–d), the dark gap is filled gradually, which shows the emergence of backscattering. The resonant backscattering, as shown by Fig. 1(e,f), only happens for certain discrete momenta. As seen from Fig. 1(e), the incident momenta for the perfect backscattering strictly conform to the resonant conditions $k_0 = n\pi/T$, where $n$ is an integer and $v_F = 1$ is assumed. In contrast, the TR symmetry broken backscattering can occur for continuously changed momenta, and more differently, it even takes place for the circularly polarized light in the off-resonant regime. Therefore, the resonant backscattering, without destroying the spin-momentum interlocking, presents a new mechanism to manipulate the TSSs. As indicated in Fig. 1(e,f), the strength of the resonant backscattering can be modulated either by changing the frequency, as shown in Fig. 1(e), or the amplitude of the gate voltage, as shown in Fig. 1(f).

Although the mechanism for the resonant backscattering induced by the gate voltage is different from that for TR symmetry broken backscattering, the underlying physics, as discussed in ref.[32], can in fact be attributed to the same origin, i.e., the nonorthogonal spins of the incident and reflection states. However, one may wonder, since, during the backscattering processes, the spin-momentum interlocking remains and the spins of the electrons are unchanged, how the TR invariant impurity scatters the surface electrons to their anti-momentum states. To further illustrate the similarities and differences between the TR invariant and TR symmetry broken scattering mechanisms, we plot the photon-dressed Dirac spectra in Fig. 2, with (a) and (c) for the time-periodic voltage, and (b) and (d) for the circularly polarized light. The off-resonant cases, i.e., $\hbar\Omega > \Lambda$, where $\Lambda$ is the bandwidth of the linear dispersion, are shown in Fig. 2(a,b), and the resonant cases with $\hbar\Omega < \Lambda$ are shown in Fig. 2(c,d). As can be seen, in the off-resonant regime, the linear spectrum and Dirac point remain for the gate voltage. However, for the circularly polarized light, the original dispersion is modified slightly, and the Kramers degeneracy due to the TR symmetry is lifted, where a band gap emerges, destroying the Dirac point. This can be easily understood from Eq. (6). By inspecting the poles of $\mathcal{G}_{\mathbf{k}}^{(0)}(\varepsilon)$, we can obtain an effective Hamiltonian for the photon-dressed spectrum

$$H_{\text{eff}} = \mathcal{H}_0 + \mathcal{H}_{-1}\mathcal{F}_1^+ + \mathcal{H}_{+1}\mathcal{F}_1^-. \tag{11}$$

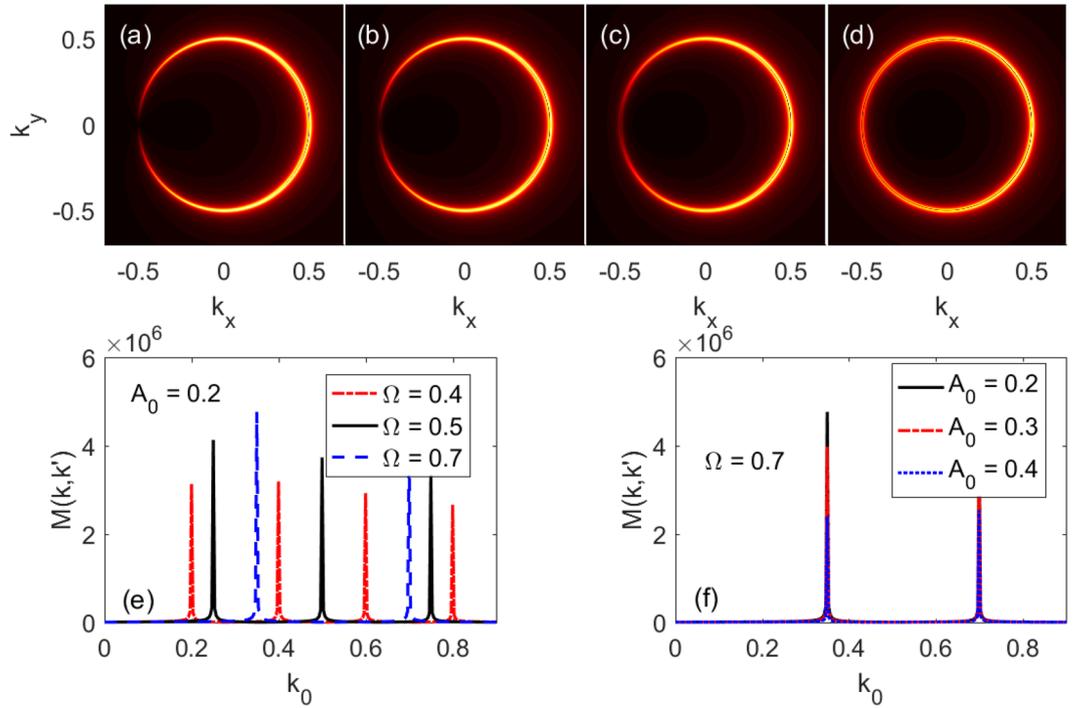

**Figure 1.** The patters of the magnetic response $M(\mathbf{k}, \mathbf{k}')$ on the $k_x - k_y$ plane, for incident momentum $\mathbf{k} = (k_0, 0)$ and scattered momentum $\mathbf{k}' = (k_x, k_y)$, for (**a**–**d**) $k_0 = \Omega/2 = 0.5$, $A_0 = 0, 0.2, 0.3, 0.5$, (**e**) $A_0 = 0.2$, $\Omega = 0.4, 0.5, 0.7$, and (**f**) $\Omega = 0.7$, $A_0 = 0.2, 0.3, 0.4$. The other parameters are chosen as $\varepsilon = \hbar v_F k_0$, $U_0 = 5$, $U_M = 0$, $A_c = 0$, $\mathbf{r}_{\text{imp}} = 0$, $\hbar = 1$ and $\Lambda = 2$.

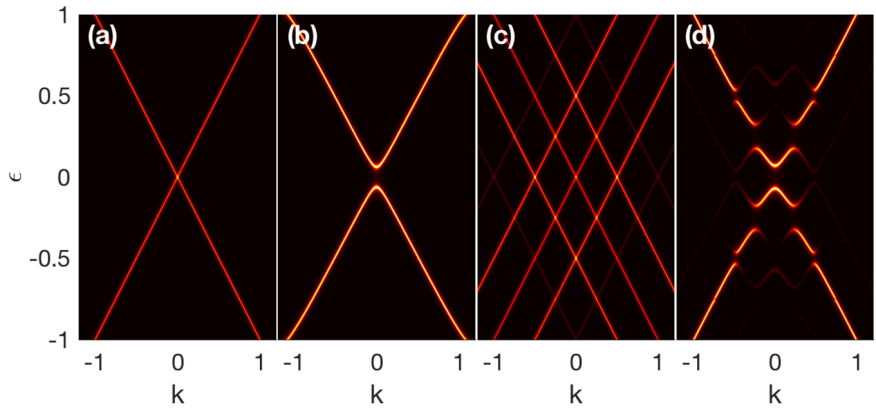

**Figure 2.** The photon-dressed spectrum for the time-periodic gate voltage with $A_c = 0$ in (**a**,**c**), and for circularly polarized light with $A_0 = 0$ in (**b**,**d**). In (**a** and **b**), we fix $\Omega = 2.5$ and set (**a**) $A_0 = 0.4$ and (**b**) $A_c = 0.4$ for off-resonant case. For resonant case, we fix $\Omega = 0.5$ and set (**c**) $A_0 = 0.38$ and (**d**) $A_c = 0.2$. The other parameters are the same as Fig. 1.

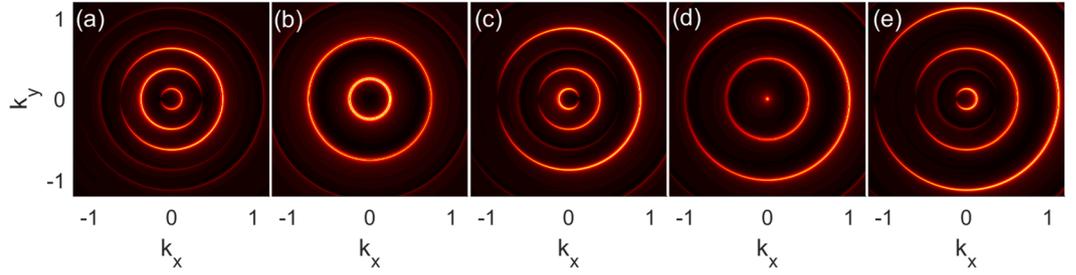

**Figure 3.** Evolution of the magnetic response function with respect to the incident momenta $k_0 = 0.125$, 0.25, 0.375, 0.5, 0.625, respectively for (**a**–**e**). Here, $A_0 = 0.3$ and $\Omega = 0.5$, corresponding to the spectrum in Fig. 2(c).

In the off-resonance regime, since the real photon absorption and emission processes are suppressed, we can approximate $\mathcal{F}_1^\pm \approx \pm \mathcal{H}_{\pm 1}/\hbar\Omega$, such that the effective Hamiltonian reduces to $H_{\text{eff}} = \mathcal{H}_0 + [\mathcal{H}_{-1}, \mathcal{H}_{+1}]/\hbar\Omega$. For the gate voltage, $\mathcal{H}_{\pm 1}$ are identical and thus commute with each other, while for the circularly polarized light, $\mathcal{H}_{\pm 1} = eA_c(\sigma_y \mp i\lambda\sigma_x)/2$, such that the commutation of $\mathcal{H}_{\pm 1}$ would generate an out-of-plane spin component for the electrons, i.e., $[\mathcal{H}_{-1}, \mathcal{H}_{+1}] = -\lambda e^2 A_c^2 \sigma_z$, which breaks the TR symmetry of the surface states. Therefore, the electrons are polarized slightly vertical to the TI surface, which makes the spins of the backscattering paired states no longer orthogonal. As a result, perfect backscattering can happen between the the paired states, even for off-resonant cases, as shown in the Fig. 1(c) in ref.[32]. However, the reduced effective Hamiltonian for the TI surfaces driven by the off-resonant gate voltage is the same as that for a static TI surface. Consequently, backscattering is still prohibited for the TSSs driven by the off-resonant gate voltage.

For resonant photon scattering, real photon absorption or emission processes will occur. As shown by Fig. 2(c,d), real photon absorption and emission split the spectrum into multi-sidebands, both for the gate voltage and circularly polarized light. The band splitting is attributed to the dynamical effective potential, i.e., the last two terms in Eq. (11), originating from the coupling between different Floquet channels. As analyzed above, the spins are invariant (modified) for the gate voltage (circularly polarized light) when the electrons absorb or emit photons. Therefore, at the crossing points of different sidebands, the states are doublet degenerate for the gate voltage. The doublet degenerate points always come in pairs and distribute symmetrically with respect to $k = 0$, as shown in Fig. 2(c). The electron spins of the paired points can be parallel or orthogonal to each other. As a result, when electrons income from these degenerate states, they can be reflected via the spin-parallel backscattering channel, such that resonant backscattering would happen. As seen from Fig. 2(c), the sidebands cross at $k = n\Omega/2 = n\pi/T$, which are exactly the resonant conditions, in accordance with the results observed in Fig. 1(e). On the contrary, for the circularly polarized light, gaps are opened up around $k = n\pi/T$, as shown in Fig. 2(d), because of the mutual coupling of the sidebands due to the slightly tip-tilted spins.

The evolution of the magnetic response function with respect to the incident momentum is plotted in Fig. 3, with the spectrum corresponding to Fig. 2(c). As compared to Fig. 1, the magnetic response function displays multiple concentric ring patterns, as more channels are opened up for a smaller frequency. When the incident momenta are away from the resonant points, as shown in Fig. 3(a,c,e), a dark gap exists in each bright ring, locating alternatively on the negative and positive $k_x$ axes, which shows the forbidden momenta for backward and forward scattering, respectively. Although the impurity can not lead to perfect backscattering $\mathbf{k} \to -\mathbf{k}$ for the off-resonant incident momenta, it can backscatter the electrons in state $(k_0, 0)$ to $(k_0 - |n|\Omega, 0)$, belonging to higher energy sidebands, by absorption of $n$-number photons. The channels $(-k_0 \pm |n|\Omega, 0)$, corresponding to the dark gaps in Fig. 3, are forbidden, because the electrons in these states have a spin direction opposite to the incident states. Therefore, from the impurity scattering patterns, one can infer the spin textures of the sidebands.

**Gate-voltage modulated local charge and spin DOS.** In the following, we study the characteristics of the gate-voltage modulated backscattering in the real space, which relate to the real-space local charge DOS $\rho(\varepsilon, \mathbf{r})$ and spin DOS $\mathbf{M}(\varepsilon, \mathbf{r})$, given by the Fourier transforms of Eqs (9) and (10), respectively. Both the physical quantities are determined by the real-space impurity perturbed Floquet Green's function, i.e.,

$$\mathcal{G}^{n,n}(\varepsilon; \mathbf{r}, \mathbf{r}') = \mathcal{G}_0^{n,n}(\varepsilon; \mathbf{r}, \mathbf{r}') + \sum_{pq} \mathcal{G}_0^{n,p}(\varepsilon; \mathbf{r}, \mathbf{r}_{\text{imp}}) \times T(\varepsilon; 0, 0) \mathcal{G}_0^{q,n}(\varepsilon; \mathbf{r}_{\text{imp}}, \mathbf{r}') \tag{12}$$

with

$$\mathcal{G}_0^{n,n}(\varepsilon; \mathbf{r}, \mathbf{r}') = \int \frac{d^2\mathbf{k}}{(2\pi)^2} e^{i\mathbf{k}\cdot(\mathbf{r}-\mathbf{r}')} \mathcal{G}_{\mathbf{k}}^{n,n}(\varepsilon) \tag{13}$$

and

$$T^{p,q}(\varepsilon; 0, 0) = V[\delta_{pq} - G^{-p,-q}(\varepsilon; 0, 0)V]^{-1}. \tag{14}$$

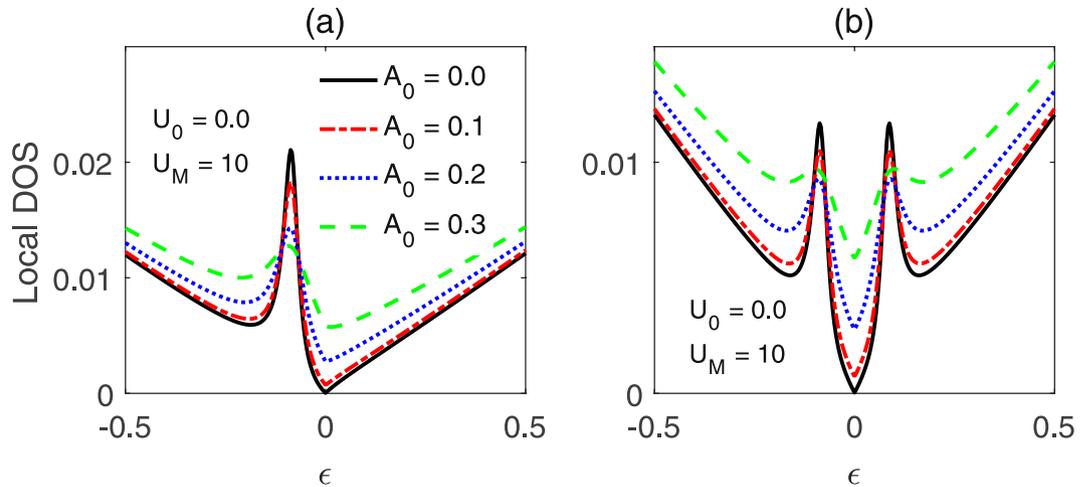

**Figure 4.** Local charge DOS for the gate-voltage-driven TI surface, with (**a**) $U_0 = 10$, $U_M = 0$, and (**b**) $U_{M,z} = 10$, $U_0 = 0$. Here, $A_0 = 0.3$, $\Omega = 1.0$ and the other parameters are the same as Fig. 1.

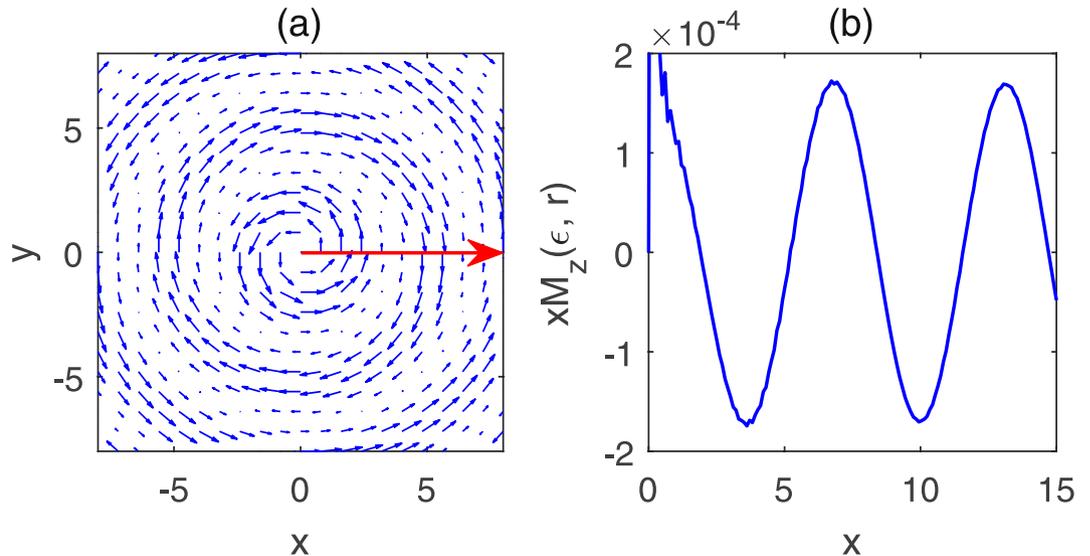

**Figure 5.** (**a**) In-plane spin textures for the impurity scattering, charaterized by $\rho M_\parallel(\varepsilon, \mathbf{r})$, where $\rho = |\mathbf{r} - \mathbf{r}_{imp}|$. (**b**) Evolution of $M_z(\varepsilon, \mathbf{r})$ along the direction marked by the red arrow in (**a**), with $U_M = 10$, $U_0 = 0$, $A_0 = 0$. The rest parameters are the same as Fig. 1.

For a static TI surface, a charged impurity can induce bound states in the local charge DOS, with a single peak either residing on the $\varepsilon < 0$ or $\varepsilon > 0$ side, which depends on the sign of $U_0$. For a magnetic impurity, a paired bound sates, symmetrically distributed with respect to the Dirac point, will be introduced into the DOS through the scattering processes. As discussed in refs[18,39], with increasing the potential energy of the impurity, the energy locations of the bound states will approach the Dirac point. The characteristics of the gate-voltage modulated local charge DOS are similar to those for a static TI surface, as shown in Fig. 4(a,b), as the spectra are similar. However, as seen in Fig. 4, shining light on the TI surface will weaken the localization for the Dirac fermions, because the electrons, by absorption or emission of photons, can hop to higher or lower energy states. The locations of the bound states, determined by the poles of the T matrix, is irrelevant to the parameters of the gate voltage, which is in contrast to the case for circularly polarized light as investigated in ref[32]. On the other hand, due to the reshuffled spectrum, as can be seen in Fig. 2(c), the DOSs around the Dirac point are raised gradually with increasing the amplitude of the light.

Due to the spin-momentum interlocking, if the spin of a surface electron deviates from its locking direction, the spin will precess around the locking direction. This can be understood from the low-energy effective Hamiltonian for the TI surface, where $\mathbf{B}_{eff} = \hbar v_F \mathbf{k}$ acts like an effective torque acting on the spin moment. According to the Landau-Lifshitz-Gilbert equation[21,49,50], i.e., $\partial_t \mathbf{s}(t) = -\mathbf{s}(t) \times \mathbf{B}_{eff}/\hbar$, the spin of a scattered electron will be finally driven to the locking direction after a long enough time, since $\mathbf{s} \parallel \mathbf{B}_{eff}$ for $\partial_t \mathbf{s}(t) = 0$. The spin precession is demonstrated in Fig. 5, where the in-plane texture exhibits a vortex structure around the impurity, with the circling directions varied periodically in the real space, accompanied with the Friedel oscillation of the out-of-plane local spin DOS. As shown in Fig. 5, when the in-plane local spin DOSs reach their maxima, the out-of-plane component vanishes, and vice versa. Together with the in-plane spin texture, the values of $M_z(\varepsilon, \mathbf{r})$ changing first from positive to negative implies that the Dirac fermions here have a left chirality. Due to the spin-orbit torque, the spin textures will decay rapidly in the real space. As indicated by the equal oscillation amplitude of $xM_z(\varepsilon, \mathbf{r})$ in Fig. 5(b), the spin texture is scaled with $1/\rho$ for a static TI surface.

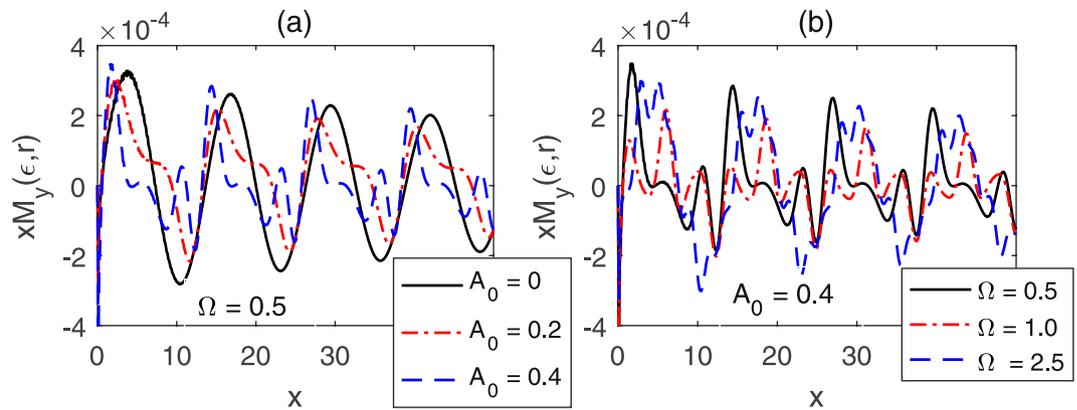

**Figure 6.** Evolution of $M_y(\varepsilon, \mathbf{r})$ along the direction marked by the red arrow in Fig. 5(a), with (**a**) $\Omega = 0.5$ and (**b**) $A_0 = 0.4$. The rest parameters are the same as Fig. 5.

As analyzed above, a nonmagnetic impurity can not induce spin textures for the gate-voltage-driven TI surface, since the spins remain unchanged during the scattering processes. However, in the presence of a time-periodic voltage, interesting spin textures will be observed for the TI surface as the electrons are scattered off a magnetic impurity. For a static TI surface, there exists only a single period in the Friedel oscillation, as can be seen from Figs 5(b) and 6(a). As discussed above, in the presence of a time-periodic gate voltage, the spectrum is split into multi-sidebands, such that new scattering channels would be opened up for the TSSs. The new scattering channels are reflected by the periods of the Friedel oscillation. Consequently, new sub-periods will emerge in the Friedel oscillation. As shown in Fig. 6(a), with increasing the amplitude of the gate voltage, the original oscillation is decorated by new sub-periods gradually. The Friedel oscillation can also be modulated by the frequency of the gate voltage. As seen in Fig. 6(b), the smaller the frequency is, the more complicated Friedel oscillation occurs, because of the more complicated reshuffled spectrum. Although the impurity scattering is dramatically modified by the gate voltage, the $1/\rho$ scale law of the spin local DOS, as indicated in Fig. 6(a,b), in which the oscillation amplitudes of $xM_y(\varepsilon, \mathbf{r})$ remains near equal for each period, persists for the gate-voltage-driven TSSs.

## Conclusion

We studied the electron scattering of the Dirac fermions on a TI surface driven by a time-periodic voltage. It is found that resonant backscattering can happen for the surface electrons incident from some discrete momenta, without breaking the TR symmetry of the TSSs. The spectrum is reshuffled in a way distinct from that for the circularly polarized light. Due to the reshuffled spectrum, new subperiods appear in the Friedel oscillation of the local spin and charge DOSs, since new channels are opened up for the electron scattering. Although the electron scattering is dramatically modified by the gate voltage, the $1/\rho$ scale law of the spin procession remains for the gate-voltage-driven TSSs.

## Acknowledgements

This work was supported by the State Key Program for Basic Researches of China under Grants No. 2015CB921202 and No. 2014CB921103 (L.S.), the National Natural Science Foundation of China under Grants No. 11674160 (L.S.), No. 11574155 (R.M.), No. 11474149 (R.S.) and a project funded by the PAPD of Jiangsu Higher Education Institutions.

## Author Contributions

M.X.D. carried out the numerical calculations. M.X.D., R.M., W.L. and L.S. analyzed the results. L.S. and D.Y.X. guided the overall project. All authors reviewed the manuscript. All authors participated in discussions and approved the submitted manuscript.

## Additional Information

**Competing Interests:** The authors declare no competing interests.

**Publisher's note:** Springer Nature remains neutral with regard to jurisdictional claims in published maps and institutional affiliations.

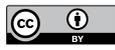**Open Access** This article is licensed under a Creative Commons Attribution 4.0 International License, which permits use, sharing, adaptation, distribution and reproduction in any medium or format, as long as you give appropriate credit to the original author(s) and the source, provide a link to the Creative Commons license, and indicate if changes were made. The images or other third party material in this article are included in the article's Creative Commons license, unless indicated otherwise in a credit line to the material. If material is not included in the article's Creative Commons license and your intended use is not permitted by statutory regulation or exceeds the permitted use, you will need to obtain permission directly from the copyright holder. To view a copy of this license, visit http://creativecommons.org/licenses/by/4.0/.

© The Author(s) 2018